\documentclass[useAMS,usenatbib]{mn2e}
\usepackage{amsmath}
\usepackage{blindtext}
\usepackage{graphicx}
\usepackage{amssymb}
\usepackage{multirow}
\usepackage{verbatim}   
\usepackage{color}     
\usepackage{subfigure} 
\usepackage{hyperref}
\usepackage{latexsym}
\usepackage{caption}
\usepackage{wrapfig}
\usepackage{soul}
\usepackage{natbib}
\usepackage{float} 

\title[Beta Function Quintessence Cosmological Parameters and Fundamental Constants I: Power and Inverse Power Law Dark Energy Potentials]{Beta Function Quintessence Cosmological Parameters and Fundamental Constants I: Power and Inverse Power Law Dark Energy Potentials}
\author[Rodger I. Thompson]{Rodger I. Thompson$^{1}$\thanks{E-mail:
rit@email.arizona.edu (RIT)}\\
$^{1}$Steward Observatory, University of Arizona, Tucson, AZ 85721, USA}

\begin{document}

\date{Accepted xxxx. Received xxxx; in original form xxxx}

\pagerange{\pageref{firstpage}--\pageref{lastpage}} \pubyear{2016}

\maketitle

\label{firstpage}

\begin{abstract}
This investigation explores using the beta function formalism to calculate analytic
solutions for the observable parameters in rolling scalar field cosmologies.  The 
beta function in this case is the derivative of the scalar $\phi$ with respect to the 
natural log of the scale factor $a$, $\beta(\phi)=\frac{d \phi}{d \ln(a)}$.  Once
the beta function is specified, modulo a boundary condition, the evolution of the
scalar $\phi$ as a function of the scale factor is completely determined.  A rolling
scalar field cosmology is defined by its action which can contain a range of physically
motivated dark energy potentials.  The beta function is chosen so that the
associated "beta potential" is an accurate, but not exact, representation of the 
appropriate dark energy model potential.  The basic concept is that the action
with the beta potential is so similar to the action with the model potential that
solutions using the beta action are accurate representations of solutions
using the model action. The beta function provides an extra equation
to calculate analytic functions of the cosmologies parameters as a function of the
scale factor that are that are not calculable using only the model action. As an
example this investigation uses a quintessence cosmology to demonstrate
the method for power and inverse power law dark energy potentials.  An interesting
result of the investigation is that the Hubble parameter H is almost completely 
insensitive to the power of the potentials and that $\Lambda$CDM is part of the
family of quintessence cosmology power law potentials with a power of zero.
\end{abstract}


\maketitle

\section{Introduction} \label{s-intro} 
The nature of dark energy is one of the key cosmological questions of our time.  A basic
component of the question is whether dark energy is static as predicted by the cosmological
constant $\Lambda$ or dynamical as predicted by rolling scalar field cosmologies. The
proper test is to determine which theory best fits the observations.  The predictions of the
cosmological constant are well known and appear to be consistent with current observations.
Ideally the predictions of scalar field cosmologies should start with the action of the 
cosmology which can accommodate various physically motivated model dark energy potentials
$V(\phi)$ where $\phi$ is the scalar field.  Unfortunately it is often mathematically difficult 
or impossible to make calculations based on the resulting action even for simple dark
energy models such as power law potentials \citep{nar17}.  This work investigates the
use of the beta formalism to provide accurate analytic equations for the evolution of
cosmological parameters as a function of the observable scale factor $a$ as opposed
to the generally unobservable scalar $\phi$.

The beta function is defined as the derivative of the scalar with respect to the natural log
of the scale factor
\begin{equation} \label{eq-beta}
\beta(\phi) \equiv \frac{d \phi}{d \ln(a)} =\phi'
\end{equation}
where the second equality notes the common cosmological practice of denoting the
derivative with respect to $\ln(a)$ with a prime.  As described in section~\ref{s-beta}
the beta function is chosen so that the resultant "beta potential" is an accurate 
representation of the model dark energy potential in the model action.  For most cases 
the action with the beta potential is so similar to the action with the model potential that
solutions using the beta action are accurate representations of solutions using the model action.
Once the form of the beta function is defined analytic solutions of the evolution 
of the cosmological parameters can be found as a function of the scalar $\phi$.  The beta 
function also provides the means to express the solutions in terms of the scale factor 
$a$ rather than the scalar $\phi$. This investigation explores the bounds of the 
parameter space where the beta 
function formalism produces solutions that deviate from the exact solution by only on 
the order of $1\%$ or less.  The primary purpose of the investigation is the 
provision of accurate, analytic functions of the evolution of the cosmological parameters
to determine which cosmologies and potentials are consistent with the observed 
universe and which must be discarded as untenable in the face of the data.  The 
functions also serve as excellent starting points for more exact numerical calculations.

The beta function formalism has its roots in a perceived correspondence between
cosmological inflation and the Quantum Field Theory renormalization group flow
equation \citep{bin15, cic17, koh17}.  In that context it is valid as the solution
for the slow evolution of a system approaching or leaving a critical (fixed) point \citep{bin15}.
Both \citet{bin15} and \citet{cic17} have considered the formalism for the late time
dark energy inflation where the critical point is in the infinite future.  The descriptions
here follow these references with particular dependence on \citet{cic17} who have
incorporated matter as well as dark energy in order to describe a real universe.  

The beta function formalism is often associated with the term universality \citep{bin15,
cic17, koh17} referring to a commonality among seemingly disparate cosmologies revealed by the
beta function formalism.  The example used in this work is too limited to fully show this but
section~\ref{ss-ha} hints at this where a common analytic function is found for the Hubble
parameter $H=\frac{\dot{a}}{a}$ which is shared by $\Lambda$CDM.  

This work concentrates on the "late time" evolution of the universe which is taken to be
the time between a scale factor of 0.1 and 1.0 corresponding to redshifts between zero
and nine.  As a demonstration of the method a quintessence cosmology is considered 
with power and inverse power law dark energy potentials.  Natural units with $\frac{8\pi G}{3}$ 
and the Planck mass equal to one are used.  A flat universe is assumed with $H_0 = 70$ 
km/sec per megaparsec. The current ratio of the dark energy density to the critical density 
$\Omega_{\phi_0}$ is set to 0.7 where $\phi_0$ is the current value of the scalar $\phi$.  
The analytic functions have $H_0$ and $\Omega_{\phi_0}$
as parameters therefore results for other choices are easily obtained.  Integer powers of
$\phi$ are taken to be $\pm(1, 2, 3, 4, 5)$ as examples but the derived functions are
valid for fractional powers as well.  The current values of the dark energy equation of 
state $w=\frac{p_{\phi}}{\rho_{\phi}}$ are taken to be $w_0=(-0.98, -0.96, -0.94, -0.92,
-0.90)$ where $p_{\phi}$ is the dark energy pressure and $\rho_{\phi}$ is the dark energy
density.  The last two values of $w_0$ are unlikely but are included to determine the 
limits of the formalism.

\section{Quintessence} \label{s-q}
Quintessence is of the most studied rolling scalar field cosmologies still standing after 
the observation of gravity waves from merging neutron stars \citep{ezq17,dur18}.  It is 
characterized by an action of the form
\begin{equation} \label{eq-act}
S=\int d^4x \sqrt{-g}[\frac{R}{2}-\frac{1}{2}g^{\mu\nu}\partial_{\mu}\partial_{\nu}\phi -V(\phi)]
+S_m
\end{equation}
where $R$ is the Ricci scalar, $g$ is the determinant of the metric $g^{\mu\nu}$, $V(\phi)$ 
is the dark energy potential, and, $S_m$ is the action of the matter fluid.  Different types
of quintessence are defined by different forms of the dark energy potential.   

The dark energy density, $\rho_{\phi}$, and pressure, $p_{\phi}$, are derived from the 
energy momentum tensor which again involves $V(\phi)$.  
\begin{equation} \label{eq-rhop}
\rho_{\phi} \equiv \frac{\dot{\phi}^2}{2}+V(\phi), \hspace{1cm} p_{\phi}  \equiv \frac{\dot{\phi}^2}{2}-V(\phi)
\end{equation}
An essential observable cosmological parameter is the dark energy equation of state
$w=\frac{p_{\phi}}{\rho_{\phi}}$.  Note that if $\dot{\phi}$ is zero then $w=-1$ for
all time as in $\Lambda$CDM.  For a quintessence cosmology \citet{nun04} give 
the dark energy equation of state as
\begin{equation} \label{eq-nun4}
w+1 =\frac{\phi'^2}{3 \Omega_{\phi}} =\frac{\beta^2(\phi)}{3 \Omega_{\phi}}
\end{equation}
where $\Omega_{\phi}$ is the ratio of the dark energy density to the critical density.
The factor $\Omega_{\phi}$ recognizes that there can be matter as well as dark energy 
in the universe so that for a flat universe with matter $\Omega_{\phi}$ is not 1 but rather 
$1-\Omega_m$ where $\Omega_m$ is the ratio of the matter density to the critical density. 
The current value of the equation of state $w_0$ is therefore a possible boundary condition
in the solution for the scalar $\phi$.

\section{The Beta Function} \label{s-beta}
The beta function is defined in eqn.~\ref{eq-beta} as the derivative of the scalar with respect to 
the natural log of the scale factor. Analytic solutions for the cosmological parameters are possible 
because the beta function provides an additional equation that determines the evolution of the 
scalar $\phi$ as a function of the scale factor.  The beta function is not an arbitrarily chosen 
relation of $\phi$.  It is directly tied to the physically relevant model dark energy potential $V(\phi)$ 
in the action.

For a given model potential $V(\phi)$, the beta function $\beta(\phi)$ is chosen so that
\begin{equation} \label{eq-bv}
V_m(\phi)=\exp\{-\int\beta(\phi)d\phi\}
\end{equation}
where $V_m(\phi)$ is the model potential rather than the full potential given
in eqn.~\ref{eq-v}.
With the proper choice of $\beta(\phi)$ any function for $V(\phi)$ can be represented, not just 
the functions considered in this investigation.  From eqn.~\ref{eq-bv} $\beta(\phi)$ is chosen
such that the integral of $\beta(\phi)$ equals  the logarithmic derivative of $V$.  The power and 
inverse power law potential beta functions are then
\begin{equation} \label{eq-betap}
\beta(\phi) = \frac{-\beta_p}{\phi}, \hspace{1cm} \beta(\phi) = \frac{\beta_i}{\phi}.
\end{equation}
where $\beta_{p,i}$ are positive numbers equal to the power.  The subscripts $p$ and $i$
are used to denote power law and inverse power law respectively.  The
scalar $\phi$ is positive for both the power and inverse power law cases.

\begin{figure}
\scalebox{.6}{\includegraphics{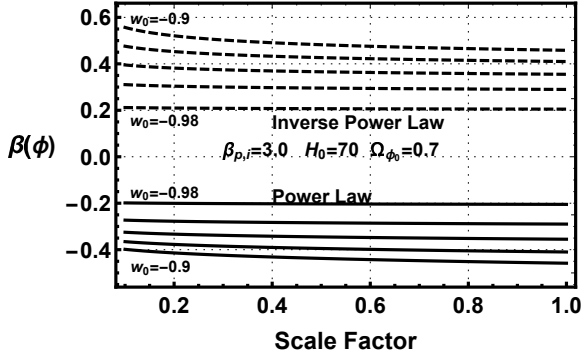}}
\caption{The evolution of the beta function $\beta(\phi)$ as a function of the scalar $a$
with $\beta_{p,i}=3$ and the five different values of $w_0$. The power law $\beta(\phi)$ 
(solid line) is negative and the inverse power law  $\beta(\phi)$ (dashed lined) is positive.}
\label{fig-betaw}
\end{figure}

\begin{figure}
\scalebox{.6}{\includegraphics{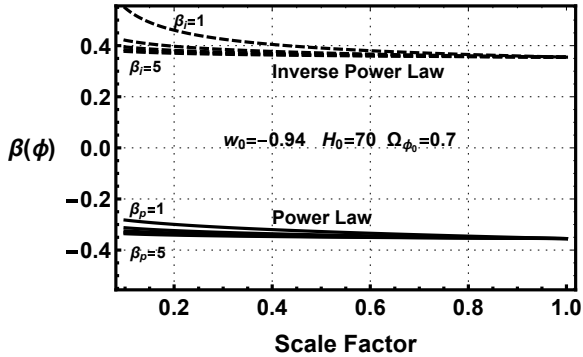}}
\caption{The evolution of the beta function $\beta(\phi)$ as a function of the scalar $a$
with $w_0=-0.94$ for the five values of $\beta_{p,i}$.}
\label{fig-betab}
\end{figure}

Figure~\ref{fig-betaw} shows the evolution of the beta functions for $\beta_{p,i}$ held
constant at 3.0 for the five different values of $w_0$.  Except where otherwise noted in
subsequent figures power law functions are denoted with a solid line and inverse power
law functions with a dashed line.  
In figure~\ref{fig-betab} $w_0=-0.94$ for all five $\beta_{p,i}$ values for the power and
inverse power law potentials.
Note that the values of $\beta(\phi)$ for a given value of $\beta_{p,i}$ are sensitive to
the value of $w_0$ but for a given value of $w_0$ the values are relatively insensitive to
$\beta_{p,i}$.  This is a pattern that occurs for many of the functions and parameters
considered here.

\section{Evolution of the Scalar} \label{s-es}
From the definition of the beta function a simple integration of eqns.~\ref{eq-betap} gives
\begin{equation} \label{eq-phi}
\phi_p(a) = \sqrt{-2 \beta_p ln(a) + \phi_0^2}, \hspace{0.5cm} \phi_i(a) = \sqrt{2 \beta_i ln(a) + \phi_0^2}
\end{equation}
where $\phi_0$ is the present day value of $\phi$.  As is evident when
$\beta(\phi)$ is used in eqn.~\ref{eq-nun4} to replace $\phi'$ the value of $\phi_0$ is related
to the current dark energy equation of state $w_0$ by 
\begin{equation} \label{eq-phio}
\phi_0 =\frac{\beta_{p.i}}{\sqrt{3 \Omega_{\phi_0}(1+w_0)}}
\end{equation}
for a quintessence cosmology where $\Omega_{\phi_0}$ is the current value
of $\Omega_{\phi}$.   Note that $\phi_0$ is the same for both the 
power and inverse power law beta functions with the same values of $\beta_{p,i}$. 

\subsection{Limitations on the Inverse Power Law Beta Function} \label{ss-bl}
At all times in the past the value of $\ln(a)$ is negative, therefore, the term in the square root
for the inverse power law in eqn.~\ref{eq-phi} becomes negative at some time in the past
limiting the range of the scale factor.  This is one justification 
for considering the power law and inverse power law beta functions as two separate cases.  
For the inverse power law case $\phi_0^2$ must be larger than $|2\beta_i \ln(a)|$ to avoid a 
negative argument.  Using eqn~\ref{eq-phio} this sets a requirement that
\begin{equation} \label{eq-brec}
2\ln(a) +\frac{\beta_i}{3\Omega_{\phi_0}(w_0+1)} >0
\end{equation}
to insure that $\phi$ is a real number. For the scale factors between 0.1 and 1 considered in
this work the constraint in eqn.~\ref{eq-brec} is  satisfied for all values of $\beta_i$ and 
$w_0$ utilized in the investigation.  For $\beta_i=1$  and $w_0=-0.9$, however, it is not satisfied
at scale factors less than 0.0925, very close to the smallest scale factor of 0.1.  As $2\beta_i\ln(a)$ 
approaches $-\phi_0^2$ the beta function evolves rapidly to large numbers making the solutions 
in this region unreliable. The increased deviation of the $\beta_i=1$ track in fig.~\ref{fig-betab} is
an indicator of the problem. A restriction that only scale factors that are at least some number 
larger than the scale factor where the argument of eqn.~\ref{eq-brec} becomes zero are 
considered reliable could be adopted.  Instead in section~\ref{ss-fit} a more physically motivated 
restrictions are imposed on the scale factors based on the accuracy of the beta potentials match 
to the model potential.  These restrictions are applied to both the power law and inverse 
power law potentials. 

\subsection{The Scalar as a Function of the Scale Factor}
\begin{figure}
\scalebox{.6}{\includegraphics{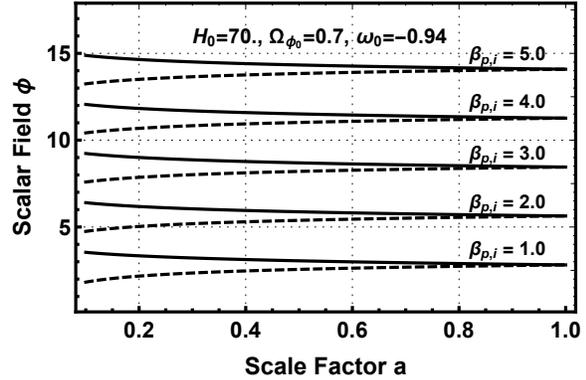}}
\caption{The evolution of the scalar field $\phi$ as a function of the scalar $a$ for the power
and inverse law beta function with $w_0 = -0.94$ for the five values of $\beta_{p,i}$. }
\label{fig-phip}
\end{figure}

\begin{figure}
\scalebox{.6}{\includegraphics{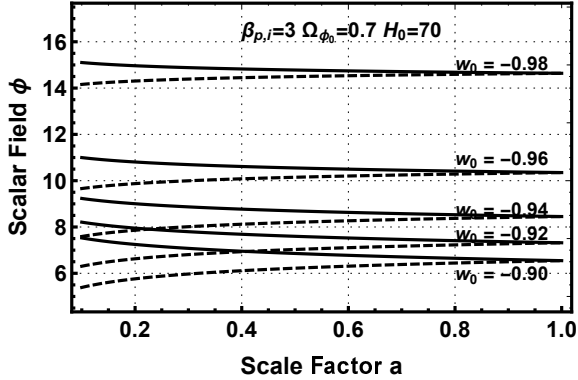}}
\caption{The evolution of the scalar field $\phi$ as a function of the scalar $a$ for the power
and inverse law beta function with $\beta_{p,i} =3.0$ and the five values of $w_0$.. The power 
law scalar (solid line) decreases to $\phi_0$ and the inverse power law scalar (dashed line) 
increases to $\phi_0$.}
\label{fig-phii}
\end{figure}

\begin{figure}
\scalebox{.6}{\includegraphics{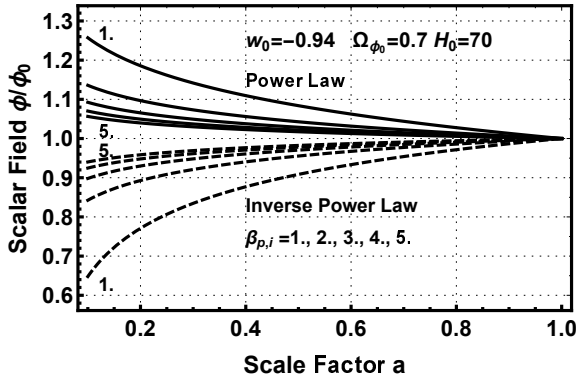}}
\caption{The evolution of the ratio of $\phi$ to $\phi_0$ with $w_0=0.96$ for the five
different values of $\beta_{p,i}$.}
\label{fig-rb}
\end{figure}

\begin{figure}
\scalebox{.6}{\includegraphics{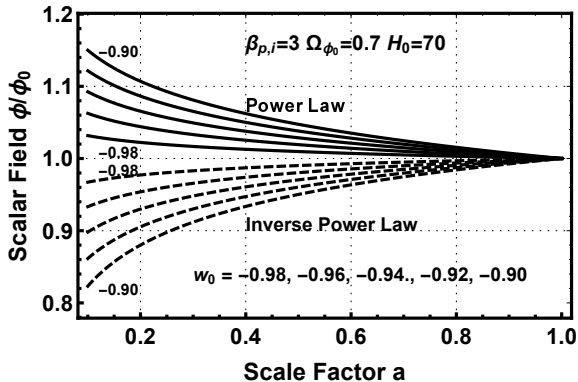}}
\caption{The evolution of the ratio of $\phi$ to $\phi_0$ with  $\beta_{p,i}=3$ for the five
different values of  $w_0$.}
\label{fig-rw}
\end{figure}

Figure~\ref{fig-phip} shows an example of the evolution of $\phi$ for both the power and inverse
power law cases for $w_0=0.94$. 
The power law scalar decreases as the scale factor increases while the inverse power law scalar 
increases with increasing $a$.  Both converge to the same value ($\phi_0$) at $a=1$.  Even though 
$\phi_0$ changes significantly with the value of $\beta_{p,i}$, the scalar $\phi$  evolves relatively 
little over $a$ between 0.1 and 1.  Figure~\ref{fig-phii} shows the evolution of the scalar with 
$\beta_{p,i} = 3.0$ and the five different values of $w_0$.
Figures~\ref{fig-rb} and~\ref{fig-rw} quantify the small variation of $\phi$ by plotting 
the ratio of $\phi$ to $\phi_0$ with $w_0 = -0.94$ in fig.~\ref{fig-rb} and for the five values 
of $w_0$ with  $\beta_{p,i} = 3$ in fig.~\ref{fig-rw}.  The figures show that the scalar 
varies by relatively little over the look back time of 13 gigayears considered in this study.  
They also show that smaller values of $\beta_{p,i}$ and larger deviations of $w_0$ from
minus one result larger changes in $\phi/\phi_0$.

\begin{figure}
\scalebox{.6}{\includegraphics{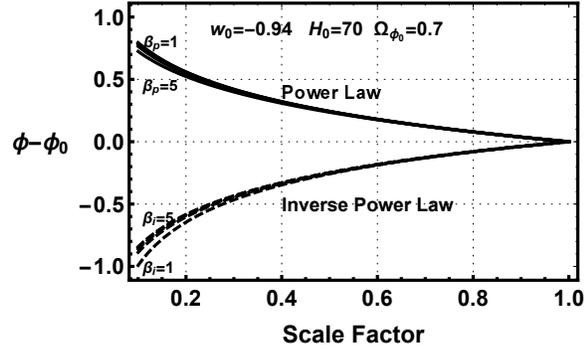}}
\caption{The evolution of  $\Delta \phi = \phi - \phi_0$ with  $w_0= -0.94$ for the five
different values of  $\beta_{p,i}$.}
\label{fig-delphi}
\end{figure}

In some cases the evolution of a parameter depends on the absolute change in the scalar
$\Delta \phi = \phi - \phi_0$ rather than the relative change in $\phi$.  Figure~\ref{fig-delphi}
shows the values of $\Delta \phi$ for the five values of $\beta_{p,i}$ for $w_0 = -0.94$.  The
value of $\Delta \phi$ is essentially independent of the value of $\beta_{p,i}$ for a given value
of $w_0$. This is a primary factor in the later conclusions that several parameters appear
insensitive to the power, $\beta_{p,i}$, of the power laws considered in this work.

\section{The Potentials} \label{s-pot}
In the beta function formalism two potentials play a prominent role.  The first is the dark
energy potential in the action $V(\phi)$ that does not depend on matter.  The second, in 
analogy with particle physics, is termed the super potential $W$ given by
\begin{equation} \label{eq-W}
W(\phi) = -2H(\phi) = -2\frac{\dot{a}}{a}
\end{equation}
Even though the Hubble parameter $H$ is the parameter of interest the development
of the method utilizes $W$ to be consistent with the literature on beta functions.
Both the potential $V(\phi)$ and the super potential $W(\phi)$ can be expressed in terms 
of $\beta(\phi)$ \citep{cic17} by
\begin{equation} \label{eq-wphi}
W(\phi) = W_0 exp\{-\frac{1}{2}\int_{\phi_0}^{\phi}\beta(x)dx\}
\end{equation}
and
\begin{equation} \label{eq-v}
	V(\phi) = \frac{3}{4} W_0^2 exp\{-\int_{\phi_0}^{\phi}\beta(x)dx\}(1-\frac{\beta^2(\phi)}{6})
\end{equation}
where $W_0$ is the current value of $W$ equal to $-2H_0$.  Note that the super potential is
always denoted as a capital $W$ and the dark energy equation of state by a lower case $w$.

The power law beta function results in simple forms of the two potentials
\begin{equation} \label{eq-wpphi}
W(\phi) = W_0(\frac{\phi}{\phi_0})^{\frac{\beta_p}{2}}
\end{equation}
and
\begin{equation} \label{eq-vpphi}
V(\phi) =  \frac{3}{4} W_0^2 (\frac{\phi}{\phi_0})^{\beta_p}(1-\frac{\beta_p^2}{6 \phi^2}) 
\end{equation}
The inverse power law also has simple forms for the potentials.
\begin{equation} \label{eq-wiphi}
W(\phi) = W_0(\frac{\phi}{\phi_0})^{-\frac{\beta_i}{2}}
\end{equation}
and
\begin{equation} \label{eq-viphi}
V(\phi) =  \frac{3}{4} W_0^2 (\frac{\phi}{\phi_0})^{-\beta_i}(1-\frac{\beta_i^2}{6 \phi^2}) 
\end{equation}
\subsection{Normalization} \label{ss-norm}
It is clear that the beta dark energy potentials have the desired power and inverse power law 
potentials multiplied by $(1-\frac{\beta_{p,i}^2}{6 \phi^2})$ which produces both an offset 
and a deviation from the model potentials.  The deviation is expected to be 
small since $\frac{\beta_{p,i}^2}{6\phi_{p,i}^2}$ is much less than one in most cases. The 
offset can be corrected by a simple normalization $(1-\frac{
\beta_{p,i}^2}{6\phi_{p,i}^2(a_n)})^{-1}$ where $a_n$ is the scale factor where the normalization
occurs.  The average deviation can be minimized by choosing a midway point such as $a_n=0.5$,
however, in this work the normalization point is $a_n=1$, the current epoch since that is where
the boundary condition is set such that $H(a=1)=H_0$.  Numerical accuracy could be increased
by normalizing piecewise at several scale factors.  A goal of this work is to create analytic 
solutions, rather than numerical tables, therefore only one normalization point is utilized.

\subsection{Accuracy of Fit} \label{ss-fit}
\begin{figure}
\scalebox{.6}{\includegraphics{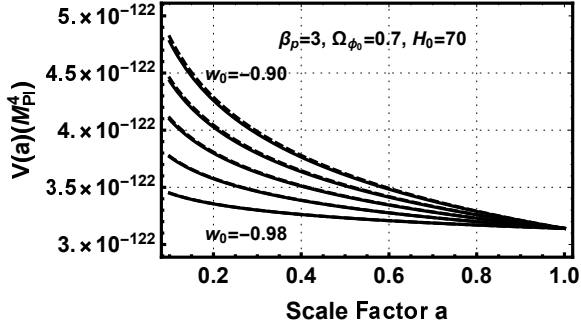}}
\caption{The solid lines show the model power law potentials with $\beta_p =3.0$ for the five 
different values of $w_0$. The dashed lines are the beta potentials for comparison.  The quality 
of the fits makes it difficult the resolve the solid lines from the dashed.  The  beta potentials are 
normalized to match the model potentials at $a=1$}
\label{fig-pdev}
\end{figure}

\begin{figure}
\scalebox{.6}{\includegraphics{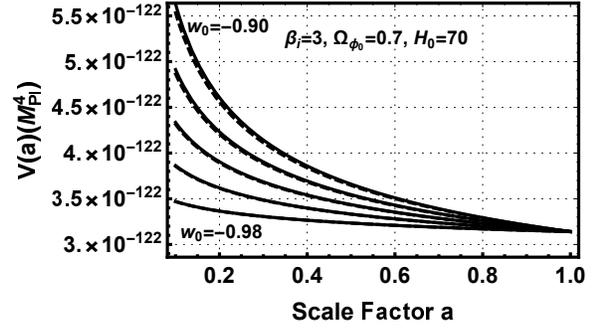}}
\caption{The same as for fig.~\ref{fig-pdev} except for the beta and model inverse power law 
potentials.}
\label{fig-idev}
\end{figure}

\begin{figure}
\scalebox{.6}{\includegraphics{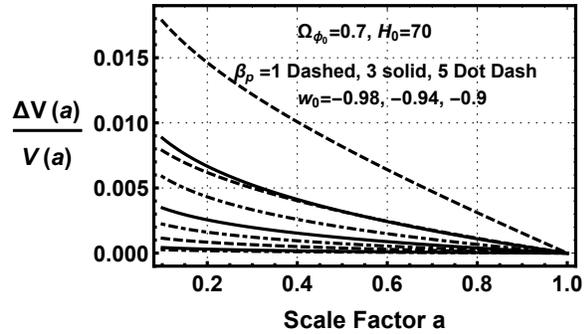}}
\caption{The fractional deviation of the beta power law potentials from the model potentials with 
$\beta_p =1.0$, dashed lines,  $\beta_p =3.0$, solid lines, and $\beta_p =5.0$, dot 
dashed lines. For each $\beta_p$ the tracks with the minimum deviation are for $w_0=-0.98$
and the tracks with the maximum deviation are for $w_0=-0.90$}
\label{fig-apfd}
\end{figure}

\begin{figure}
\scalebox{.6}{\includegraphics{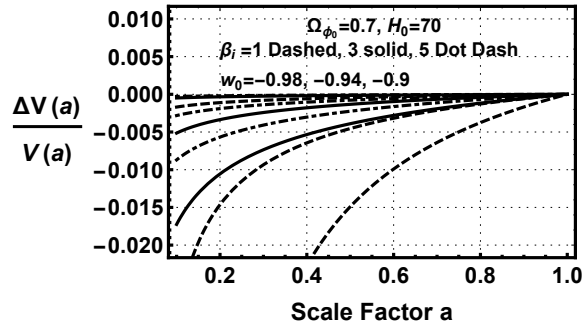}}
\caption{The same as for fig.~\ref{fig-apfd} except for the inverse power law potentials.
For $\beta_i=1$ only the $w_0=-0.98$ track is shown.}
\label{fig-aifd}
\end{figure}

The cosmological parameters derived by the beta function formalism are only useful if the
beta potentials accurately represent the model potentials.  Figures~\ref{fig-pdev} and~\ref{fig-idev}
show the evolution of the power and inverse power law potentials respectively.  In contrast to
previous figures the solid lines are the model potentials and the dashed lines are the beta 
potentials.  The value of $\beta_{p,i}$ is set to 3.0.  The beta potentials are an excellent match 
to the model potentials for the parameters in the figure.  The matches improve as $w_0$ 
approaches minus one. For a given value of $\beta_{p,i}$ the inverse power law potentials 
have about $10\%$ more evolution than the power law potentials.
Figures~\ref{fig-apfd} and~\ref{fig-aifd} show the fractional deviation of the beta potentials 
from the model potentials to quantify the deviations of the beta potentials from the model
potentials.   The $\beta_{p,i}$ values equal to 1, 3 and 5 and with $w_0$ values equal to -0.98, 
-0.94 and -0.9 are chosen to show the extremes without excessive overlap of tracks in the
figures.
The power law beta potentials are quantitatively good matches to model potentials with
the fit improving as $\beta_p$ increases and as $w_0$ decreases toward minus one.  Only
the $\beta_p = 1$ with $w_0=-0.90$ case exceeds a fractional deviation of $1\%$ and then
only at scale factors less than 0.4.  The inverse power law beta potentials show the same trends
but are less well behaved.  It is clear that for low $\beta_i$ values and large deviations of
$w_0$ from minus one some of the beta potentials deviate from the model potentials by much
more than $1\%$.

In this investigation the conservative limit of no more than $1\%$ deviation of the beta potential
from the model potential is adopted.  Tables~\ref{tab-pval} and~\ref{tab-ival} indicate the
minimum value of the scale factor for a given pair $\beta_{p,i}$ and $w_0$ where the beta
potential is within $1\%$ of the model potential.  Entries with a v indicate the $1\%$  limitation
is satisfied for all scale factors between 0.1 and 1.0 Subsequent figures adhere to this limitation.  
\begin{table}
\begin{tabular}{cccccc}
\hline
 &  &  & $w_0$  & & \\
\hline
$\beta_p$ & $-0.98$  & $-0.96$ & $-0.94$  & $-0.92$ & $-0.90$\\
\hline
1 & v &  v & v & 0.2 & 0.4\\
2 & v &  v & v & v & 0.15\\
3 & v &  v & v & v & v \\ 
4& v &  v & v & v & v \\
5& v &  v & v & v & v \\
\hline
\end{tabular}
\caption{Valid values of the scale factor for the power law beta potentials.  The scale
factor must be greater than the entered value for the given value of $\beta_p$ and
$w_0$. An entry of v indicates that all scale factors between 0.1 and 1 are valid.} \label{tab-pval}
\end{table}

\begin{table}
\begin{tabular}{cccccc}
\hline
 &  &  & $w_0$  & & \\
\hline
$\beta_i$ & $-0.98$  & $-0.96$ & $-0.94$  & $-0.92$ & $-0.90$\\
\hline
1 & v &  v & 0.28 & 0.47 & 0.6\\
2 & v &  v & v & 0.21 & 0.37\\
3 & v &  v & v & v & 0.20\\ 
4& v &  v & v & v & 0.12\\
5& v &  v & v & v & v \\
\hline
\end{tabular}
\caption{The same as table~\ref{tab-pval} for the inverse power law
potentials.} \label{tab-ival}
\end{table}

The range of $w_0$ values for this investigation was extended past -0.94 to test
the limits of the validity of the method.  For the power law beta functions the only
cases that are not valid over all scale factors are for $w_0$ values of -0.98 and -0.9
with $\beta_p$ values of 1 and 2.  The minimum $a$ value of $\beta_p=2$ and 
$w_0=-0.9$ is 0.15, therefore, most of the range of the scale factor is valid.  For the 
inverse power law case only the $\beta_i=1$ with $w_0=-0.94$ has a limitation on
the scale factor for the three values of $w_0$ nearest minus one.  This leads to the
conclusion the beta function formalism is a useful method for power and inverse
power law dark energy potentials within the expected values of $w_0$.  Caution,
however, must be exercised for $w_0$ values further from minus one than -0.94
as is shown in the tables.  It is clear from tables~\ref{tab-pval} and~\ref{tab-ival}
that as the value of $\beta_{p,i}$ approaches one the solutions for the beta potentials
deviate from the model potentials by more than $1\%$ over a larger fraction of the
scale factors under consideration.  Except for the special case of $\beta_{p,i}=0$,
$Lambda$CDM values of $\beta_{p,i}<1$ are considered unreliable and are not
considered in the investigation.

\section{Adding Matter to the Universe} \label{s-mat}
A real universe includes matter as well as dark energy.  The explicit inclusion of matter is 
discussed in \cite{cic17} and is the basis for this work.  As before there is no attempt to 
rederive the work presented there except where it is useful for clarity.  The purpose of this work 
is useful analytic models for comparison with observation rather than a theoretical extension of 
previous work.  Matter is represented by the $S_m$ term the action, eqn.~\ref{eq-act}. 

\subsection{The Matter Density} \label{ss-rhom}
The matter density $\rho_m$ follows the mass continuity equation
\begin{equation} \label{eq-mc}
\dot{\rho_m}=\rho_{m,\phi}\dot{\phi}=-3H \rho_m
\end{equation}
In keeping with the notation of \citep{bin15} and \citep{cic17} the subscript ,$\phi$ indicates
the derivative with respect to $\phi$.   This leads to the equations
\begin{equation} \label{eq-bm}
\frac{\rho_{m,\phi}}{\rho_m} = -3\frac{H}{\dot{\phi}}=-\frac{3}{\beta(\phi)}
\end{equation}
Integrating the logarithmic derivative in eqn.~\ref{eq-bm} yields the  equation for $\rho_m(\phi)$
\begin{equation} \label{eq-rhomphi}
\rho_m(\phi)=\rho_{m0}\exp(-3\int_{\phi_0}^{\phi} \frac{d \phi}{\beta(\phi)})
\end{equation}
Different beta functions produce different functions for $\rho_m$ as a function of $\phi$.
The emphasis in this work, however, is expressing the cosmological parameters as a function of the
observable scale factor $a$ rather than the unobservable scalar $\phi$.  From the definition
of $\beta(\phi)$ in eqn.~\ref{eq-beta} eqn.~\ref{eq-rhomphi} becomes
\begin{equation} \label{eq-rhoma}
\rho_m(a)=\rho_{m0}\exp(-3\int_1^a d\ln(a) )= \rho_{m0}a^{-3}
\end{equation}
as expected, independent of $\beta(\phi)$.

\subsection{The Super Potential $W$ with Mass} \label{s-spm}
The Einstein equations with mass become
\begin{equation} \label{eq-em1}
H^2 = \frac{\rho_m + \rho_{\phi}}{3}
\end{equation}

\begin{equation} \label{eq-em2}
-2\dot{H} = \rho_m + \rho_{\phi} + p_{\phi}
\end{equation}

\cite{cic17} show that the inclusion of matter results in differential equation for $W$ of the form
\begin{equation} \label{eq-difw}
WW_{,\phi} + \frac{1}{2} \beta W^2 = -2\frac{\rho_m}{\beta}
\end{equation}
For the power law beta function $\beta(\phi) = -\frac{\beta_p}{\phi}$ eqn.~\ref{eq-difw} becomes
\begin{equation} \label{eq-bcw}
WW_{,\phi} - \frac{1}{2} \frac{\beta_p}{\phi} W^2 = -2\rho_m \frac{\phi}{\beta_p}
\end{equation}
Equation~\ref{eq-bcw} is solved by multiplying it by an integrating factor that makes the left hand side an 
exact differential and the right hand side an integral that can be solved preferably analytically or by numerical
integration.  The integrating factor for the power law beta function is $\phi^{-\beta_p}$.  The equation then reads
\begin{equation} \label{eq-if}
\frac{d}{d \phi}(\frac{1}{2}W^2 \phi^{-\beta_p})=2\rho_m(\phi) \frac{\phi^{1-\beta_p}}{\beta_p}
\end{equation}
which  is a general equation for any positive value of $\beta_p$.

The derivation of the super potential deviates from the discussion of \cite{cic17} at this point
to derive $W(a)$ rather than $W(\phi)$ since the goal is observable predictions.  Substituting
eqn.~\ref{eq-rhoma} into eqn.~\ref{eq-if} results in
\begin{equation} \label{eq-wa}
\mid_{\phi_0}^{\phi}W^2 \phi^{-\beta_p} = 4 \frac{\rho_{m_0}}{\beta_p}\int_{\phi_0}^{\phi}\phi^{1-\beta_p}a^{-3}d\phi
\end{equation}
Using $d\phi=-\beta_p(-2\beta_p\ln(a)+\phi_0^2)^{-1/2}\frac{da}{a}$ gives
\begin{equation} \label{eq-iwa}
\mid_{\phi_0}^{\phi}W^2 \phi^{-\beta_p} = -4\rho_{m_0}\int_{1}^{a}x^{-4}(-2\beta_p\ln(x)+\phi_0^2)^{-\frac{\beta_p}{2}}dx
\end{equation}
Equation~\ref{eq-iwa} can also be written as
\begin{equation} \label{eq-phiwa}
\mid_{\phi_0}^{\phi}W^2 = -4\rho_{m_0}\phi^{\beta_p}\int_{1}^{a}x^{-4}\phi^{-\beta_p}dx
\end{equation}
Since $\phi(a) =(-2\beta_p\ln(a)+\phi_0^2)^{1/2}$ the super potential as a function of $a$ is
\begin{align} \label{eq-war}
W(a) = \{-4\rho_{m_0}(-2\beta_p\ln(a)+\phi_0^2)^{\frac{\beta_p}{2}} \nonumber \\
 \int_{1}^{a}x^{-4}(-2\beta_p\ln(x)+\phi_0^2)^{\frac{-\beta_p}{2}}dx+W_0^2(\frac{\phi(a)}{\phi_0})^{\beta_p}\}^{1/2}
\end{align}

The integral in eqn.~\ref{eq-war} is solved by two changes of variable.  The first change is to
let $z=(-2\beta_p \ln(a) + \phi_0^2)$ which yields
\begin{equation} \label{eq-war1}
-(\frac{1}{2\beta_p}) \exp(-\frac{3\phi_0^2}{2\beta_p})\int z^{-\frac{\beta_p}{2}} \exp(\frac{3z}{2\beta_p})dz
\end{equation}
The second change of variable is $y=-\frac{3z}{2\beta_p}$ which produces the integral
\begin{equation} \label{eq-war2}
-\frac{1}{3}(-\frac{2\beta_p}{3})^{-\frac{\beta_p}{2}}\exp(-\frac{3\phi_0^2}{2\beta_p})\int y^{-\frac{\beta_p}{2}}\exp(-y)dy
\end{equation}
The integral in $y$ in eqn.~\ref{eq-war2} is the incomplete Gamma function $\Gamma(1-\frac{\beta_p}{2},3 \ln(a)-\frac{3\phi_0^2}{2\beta_p})$.  The formal solution for the super
potential in terms of the scale factor is
\begin{align} \label{eq-awa}
W_p(a)=-\{-\frac{4\rho_{m_0}}{3}(-\frac{2\beta_p}{3})^{-\frac{\beta_p}{2}}\exp(-\frac{3\phi_0^2}{2\beta_p})(\phi_p(a))^{\beta_p}\nonumber \\
\{ \Gamma(1-\frac{\beta_p}{2},3 \ln(a)-\frac{3\phi_0^2}{2\beta_p}) - \Gamma(1-\frac{\beta_p}{2},-\frac{3\phi_0^2}{2\beta_p}) \}\nonumber  \\
 +W_0^2(\frac{\phi_p(a)}{\phi_0})^{\beta_p}\}^{1/2}
\end{align}
The negative square root is chosen since $W(a)$ is a negative quantity.

The solution for $W(a)$ in the inverse power law case is very similar to the power law.  The
integrating factor is $\phi^{\beta_i}$ rather than $\phi^{-\beta_p}$. The equivalent to
eqn.~\ref{eq-phiwa} is
\begin{equation} \label{eq-phiwai}
\mid_{\phi_0}^{\phi}W^2 = -4\rho_{m_0}\phi^{-\beta_p}\int_{1}^{a}x^{-4}\phi^{\beta_p}dx
\end{equation}
and  the formal solution for $W(a)$ for the inverse power law case is
\begin{align} \label{eq-awai}
W_i(a)=-\{-\frac{4\rho_{m_0}}{3}(\frac{2\beta_i}{3})^{-\frac{\beta_p}{2}}\exp(\frac{3\phi_0^2}{2\beta_i})(\phi_i(a))^{-\beta_i}\nonumber \\
\{ \Gamma(1+\frac{\beta_i}{2},3 \ln(a)+\frac{3\phi_0^2}{2\beta_i}) - \Gamma(1+\frac{\beta_i}{2},\frac{3\phi_0^2}{2\beta_i}) \}\nonumber  \\
\noindent +W_0^2(\frac{\phi_0}{\phi_i(a)})^{\beta_i}\}^{1/2}
\end{align}

\section{The Evolution of Cosmological Parameters} \label{s-ecp}
Establishing the analytic functions for the super potential $W$ as a function of the scale
factor $a$ provides the means for calculating the evolution of cosmological parameters.
It is obvious from its definition (eqn.~\ref{eq-W}) that super potential determines the Hubble
parameter.  Normally the discussion of the cosmological parameters would center on the Hubble
parameter but the super potential is again used here to be consistent with existing literature on 
the beta function formalism.

In the following the parameters are presented as a function of $\phi_{p,i}(a)$ with 
eqn.~\ref{eq-phi} providing the proper equations for the scalar $\phi$ as a function 
of the scale factor $a$.  This convention is adopted to preserve the dependence of
the parameters on the scalar $\phi$ while providing the means to calculate the parameters
as a function of the scale factor $a$.  An exception to this convention is the matter density
where eqn.~\ref{eq-rhoma} explicitly show that the density varies as $a^{-3}$.  Although
it should be obvious from the context the scalar will be written as $\phi_p(a)$ for the 
power law and $\phi_i(a)$ for the inverse power law but the current value of $\phi$ will
still be written as $\phi_0$ since it is the same for both cases.

\subsection{The Evolution of the Hubble Factor and the Onset of Acceleration} \label{ss-ha}
\begin{figure}
\scalebox{.6}{\includegraphics{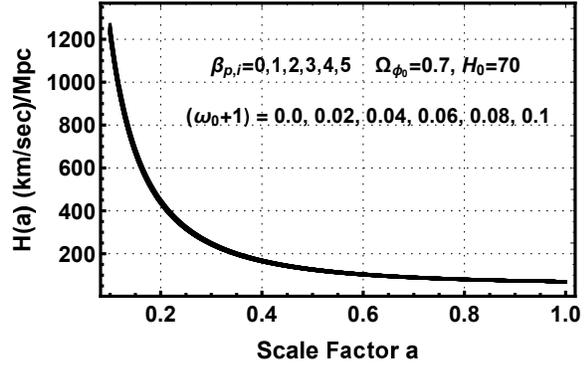}} 
\caption{The evolution of $H(a)$ for the power (solid line) and inverse (dashed line) power law 
potentials as well as $\Lambda$CDM. The inverse power law Hubble function lies very slightly 
above the power law Hubble function but the difference is not resolvable at the resolution of 
the figure.  The difference between $\Lambda$CDM and the quintessence plots 
is also not resolvable.}
\label{fig-h}
\end{figure}

Two observable quantities are the evolution of the Hubble factor $H(a)$ and the onset
of the acceleration of the expansion of the universe.  Since $H(a) = -\frac{W(a)}{2}$ 
eqns.~\ref{eq-awa} and~\ref{eq-awai} specify the evolution of the Hubble factor for the 
power and inverse power law potentials.  Figure~\ref{fig-h} shows the evolution of $H(a)$
for all of the cases considered in this study \emph{including $\Lambda$CDM}.  All of the
solutions plotted in fig.~\ref{fig-h} conform to the limits on $a$ in tables~\ref{tab-pval}
and~\ref{tab-ival}.
Remarkably the solutions for both the power and inverse power law as well as $\Lambda$CDM 
all overlap each other at the resolution of fig.~\ref{fig-h}, making $H(a)$ insensitive to 
either the power of the potential or the current value of the dark energy equation of state,
including $w_0=-1$, for the cases considered here.

\subsubsection{The insensitivity of $H(a)$ to the Potential and $w_0$} \label{sss-hins}
At first glance the insensitivity of the Hubble value $H(a)$ to the power of the potential
and the value of $w_0$ seems remarkable but further
examination shows that it is due to a combination of factors.  The first is that all solutions
must have the same initial value $H_0$ which is set by observation, independent of 
$w_0$.  A second factor is that at early times when the evolution is matter
dominated the common $\rho_m = \rho_{m_0}a^{-3}$ term makes the evolution the same
for all cases.  Thirdly the last term in both eqns.~\ref{eq-awa} and~\ref{eq-awai} is
proportional to either $\frac{\phi}{\phi_0}$ for the power law or  $\frac{\phi_0}{\phi}$
for the inverse power law.  Examination of fig.~\ref{fig-phip} shows that the power and 
inverse power law scalars are decreasing and increasing respective with $a$ making both
late time evolutions decreasing with increasing $a$. Finally examination of eqns.~\ref{eq-phiwa}
and~\ref{eq-phiwai} reveal that the integrals are multiplied by opposing positive and negative
powers of $\phi$ inside and outside of the integral.  Since the change in $\phi$
is small the positive and negative powers of $\phi$ effectively cancel each other.

\subsubsection{A simple common equation for $H(a)$} \label{sss-hcom}
Equations.~\ref{eq-phiwa} and~\ref{eq-phiwai} suggest that the integral over $x$ with 
the $\phi$ term held constant may be an excellent approximation for describing $H(a)$.  
That approximation is given by
\begin{equation} \label{eq-hcomp}
H(a)=-\frac{1}{2}\sqrt{\frac{4}{3}\rho_{m_0}(a^{-3}-1)+W_0^2(\frac{\phi(a)}{\phi_0})^{\beta_p}}
\end{equation}
for the power law case and
\begin{equation} \label{eq-hcomi}
H(a)=-\frac{1}{2}\sqrt{\frac{4}{3}\rho_{m_0}(a^{-3}-1)+W_0^2(\frac{\phi_0}{\phi(a)})^{\beta_i}}
\end{equation}
for the inverse power law case. Equation~\ref{eq-phi} provides the appropriate $\phi(a)$. 
Equations~\ref{eq-hcomp} and~\ref{eq-hcomi} give $H(a)$ solutions that are indistinguishable 
from the suite of solutions shown in figure~\ref{fig-h} at the resolution of the plot.  It is interesting 
to note that $\Lambda$CDM is the $\beta_{p,i}=0$ case for either equation making $\Lambda$CDM
a member of the family of solutions.  This is an indication of the universality of the formalism.

\subsubsection{The onset of acceleration} \label{sss-ac}
\begin{figure}
\scalebox{.6}{\includegraphics{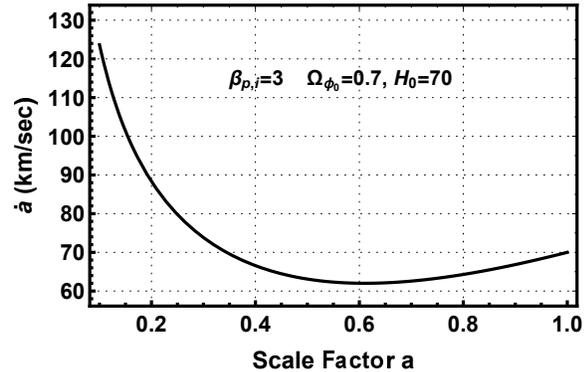}} 
\caption{The figure shows the evolution of $\dot{a} = aH(a)$ versus the scale factor
$a$ for both the power and inverse power law potentials.  The  power law potential 
track is slightly below the inverse power law track but the difference is indistinguishable
at the scale of the figure.}
\label{fig-hac}
\end{figure}

In a universe with mass the onset of the acceleration of the expansion is delayed until
the matter density is low enough that dark energy begins to dominate.  The onset of
acceleration is marked by an increase in the expansion rate $\dot{a} = aH(a)$.  
Figure~\ref{fig-hac} shows the track of $\dot{a}$ versus $a$.
The acceleration begins at a scale factor of $\approx0.6$ $(z\approx0.7)$ which is consistent 
with observations eg. \citep{avs14, avs17}.  Given the insensitivity of H(a) to $\beta_{p,i}$ 
and $w_0$ the only adjustable parameters are $H_0$ and $\rho_{m_0}$ which are set by 
observation.

\subsubsection{Comparison with Observations} \label{sss-hob}
\begin{figure}
\includegraphics[width=7.6cm]{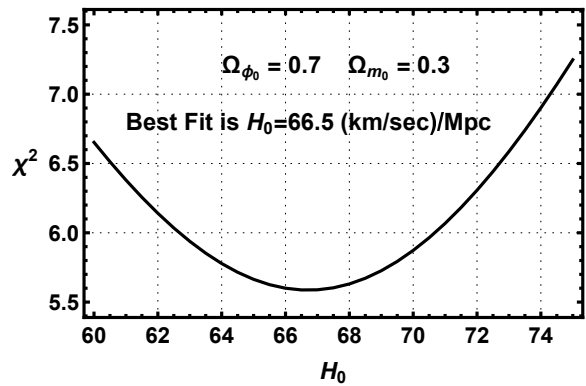}
\caption{The values of $\chi^2$ versus $H_0$ for the \citet{jes17} data set showing
a best fit of $H_0=66.5$ as the best fit for this particular data set.} 
\label{fig-hchi}
\end{figure}

We have shown that the Hubble factor $H(a)$ is remarkably insensitive to either the
power of the potential, $\beta_{p,i}$ or $w_0$ and is identical to the $\Lambda$CDM 
$H(a)$ solution. This
makes the Hubble factor a poor parameter for discriminating between static and dynamical
dark energy.  It, however, offers an excellent opportunity for determining $H_0$ for both
cosmologies.  The recently compiled $H(a)$ observations by \citet{jes17} provides an 
example of such a measurement.  Using eqn.~\ref{eq-hcomp} as the model with
$H_0$ as the only variable a chi squared analysis determined that the most likely value 
of $H_0$ for the example data set is $H_0 = 66.5$ (km/sec)/Mpc. Figure~\ref{fig-hchi} shows
the run of $\chi^2$ versus $H_0$. This is not a result, just an example for the particular data set.

\begin{figure}
\includegraphics[width=7.6cm]{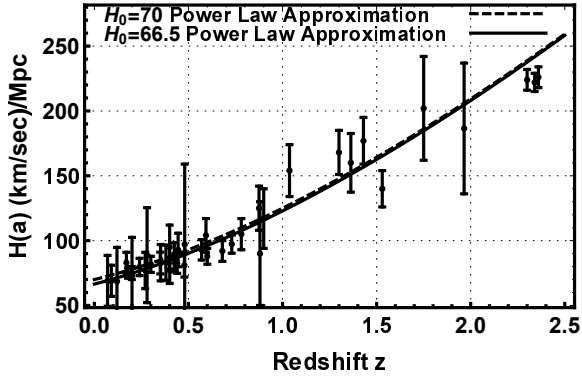}
\caption{The $H_0 = 66.5$ best fit and the $H_0=70$ proposal example fit to the
$H(z)$ data set.} 
\label{fig-hfit}
\end{figure}

Figure~\ref{fig-hfit} shows the example $H_0=70$ and the best fit $H_0=66.5$ 
$H(a)$ evolution superimposed on the \citet{jes17} data set. The dashed curve for the
$H_0=70$ case is just barely resolved above the solid line. The minimum chi square of 
about 5.6 is not a high quality measurement but is probably consistent with the scatter in
the data set providing evidence that the beta function calculations have more than sufficient 
accuracy for comparison with observations.

\subsection{The Dark Energy Equation of State} \label{ss-deos}
One of the most important observable cosmological parameters is the dark energy equation of
state $w$.  The static $\Lambda$CDM cosmology predicts that $w$ equals minus one for all time
whereas dynamical cosmologies predict values deviant from minus one.  It should be noted that
$w$ need not vary to produce dynamical cosmological parameters, it just needs to be different
from minus one.  Section~\ref{s-fc} on fundamental constants is an example of such a case.
From \citet{cic17} the dark energy equation of state is given by
\begin{equation} \label{eq-wden}
1+w(\phi) = \frac{\beta^2}{3}(1-\frac{4\rho_{m_0}a^{-3}}{3W^2})^{-1}
=\frac{\beta^2}{3}(1-\Omega_m)^{-1}=\frac{\beta^2(\phi)}{3\Omega_{\phi}}
\end{equation}
for a flat universe where the terms after the first equality are provided by the author for 
clarity.  The second equality shows that $1+w(\phi)$ is proportional to $(1-\Omega_m)^{-1}$.
In the matter dominated era $\Omega_m$ approaches one making $(1-\Omega_m)^{-1}$
very susceptible to small errors in $\Omega_m$.  For this reason the analytic solutions for
$(1+w)$ employ eqns.~\ref{eq-awa} and~\ref{eq-awai} for $W(a)$ rather than the 
approximations for $W(a)$ and $H(a)$ in eqns.~\ref{eq-hcomp} and~\ref{eq-hcomi}.
In terms of the scale factor $a$ the dark energy equation of state $w(a)$ is given by
\begin{equation} \label{eq-wdea}
1+w(a) = \frac{\beta_{p,i}^2}{3 \phi^2_{p,i}(a)}(1-\frac{4\rho_{m_0}}{3a^3W^2_{p,i}(a)})^{-1}
\end{equation}
In eqn.~\ref{eq-wdea} $\phi_{p,i}(a)$ is given by eqns.~\ref{eq-phi} and $W_{p,i}(a)$
by eqn.~\ref{eq-awa} and~\ref{eq-awai}.

\begin{figure}
\scalebox{.6}{\includegraphics{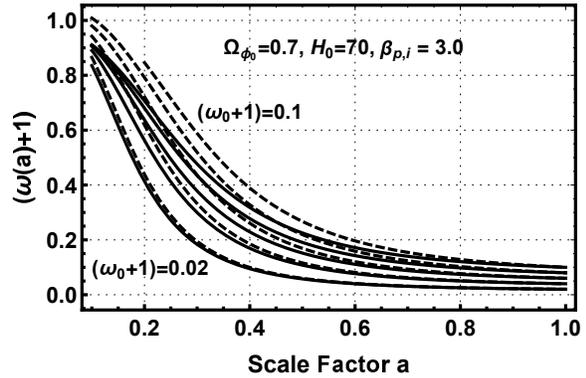}}  
\caption{The dark energy equation of state plus one with $\beta_{p,i}$ held constant at 3.0
for the five different values of $w_0$. The power law cases are plotted with a solid line and 
the inverse power law cases with a dashed line.}
\label{fig-ww}
\end{figure}

Figure~\ref{fig-ww} shows the evolution of $1+w(a)$ for $\beta_{p,i}=3.0$ with the
five values of $w_0$. 
As expected the evolution of $(1+w(a))$ is slowly freezing toward $w_0$ for scale factors
larger than 0.5 while there is significant evolution for scale factors smaller than 0.5.  At increased
deviations of $w_0$ from minus one the inverse power law cases increasingly deviates from the
power law cases.

\begin{figure}
\scalebox{.6}{\includegraphics{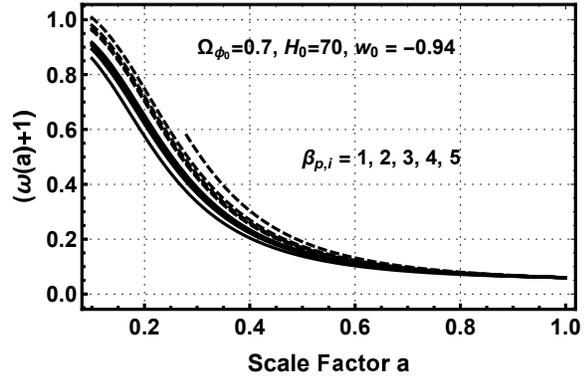}}  
\caption{The dark energy equation of state plus one with $w_0 = -0.94$ for the five different
values of $\beta_{p,i}$.  For both the power law (solid) and the inverse power law (dashed) cases
the degree of evolution decreases slightly with increasing $\beta_{p,i}$.  All of the inverse
power law tracks lie above the power law tracks.}
\label{fig-wb}
\end{figure}

Figure~\ref{fig-wb} shows the evolution of $1+w(a)$ with $w_0=-0.94$ for the five different
values of $\beta_{p,i}$.
As with the other cosmological parameters the power $\beta_{p,i}$ of the power and inverse
power law potentials has only a small effect on the evolution of $w(a)$.  Since the value of 
$w_0$ was held constant all of the cases have the same present day value of $w(a)$.

\subsection{The Fundamental Constants} \label{s-fc}
Beyond the cosmological parameters the fundamental constants provide important and 
generally under utilized information to discriminate between static and dynamic dark
energy.  Fundamental constants are pure numbers with no dimensions and therefore
invariant to the system of units.  The constants considered here are the proton to electron
mass ratio, $\mu$, and the fine structure constant, $\alpha$.  The standard model, with
the cosmological constant as dark energy, predicts that the fundamental constants are
temporally and spatially invariant.  Quintessence and most other rolling scalar field
cosmologies predict a temporal variation of the constants that is proportional to the
deviation of $w$ from minus one \citep{cal11,thm12}.  This connection occurs because
the scalar $\phi$ that provides dark energy also interacts with other sectors beyond
the gravitational sector.

 In the absence of special and  finely tuned symmetries it is very difficult to restrict 
a scalar field that interacts with gravity from interacting with the weak, electromagnetic and 
strong sectors as well eg. \citep{car98, ave06}. In this scenario the same field $\phi$ that 
serves as dark energy also produces changes in the fundamental constants and particle 
physics parameters through interactions in sectors other than gravity.  The values
of the fundamental constants such as the proton to electron mass ratio $\mu$ and the
fine structure constant $\alpha$ are set by the values of the particle physics parameters 
such as the Quantum Chromodynamic Scale, $\Lambda_{QCD}$, the Higgs Vacuum Expectation 
Value, $\nu$, and the Yukawa couplings, $h$.  The scalar interacting with these particle
physics parameters produces changes in the fundamental constants \citep{coc07,thm17}.

The coupling of the scalar field to $\alpha$ and $\mu$ is given by the simple relation
\citep{nun04}
\begin{equation} \label{eq-dx}
\frac{\Delta c}{c} = \zeta_c(\phi - \phi_0), \hspace{0.5cm} c=\alpha,\mu
\end{equation}
The coupling to the constant $c$ is $\zeta_c$, which may be either positive or negative.  The 
linear dependence of the variance of the constants on $\phi$ can be thought of as the first
term in a Taylor series expansion of a more complicated coupling.  Since the limits on 
observed changes in the constants are on the order of $10^{-6}$ or less the linear dependence
is a good approximation.  Although $\zeta_{\mu}$ is written as a single term
it is actually a combination of the individual couplings to the QCD scale, the Higgs VeV and
the Yukawa couplings as discussed above, in \citet{thm17}, and at the end of  
Section~\ref{ss-muob}.  It is clear from eqn.~\ref{eq-dx} that once the beta function
is defined and the boundary condition selected  the evolution of the fundamental
constants is completely defined.  This is one of the significant advantages of the beta 
function formalism.  Using the connection between $w$ and $\phi$ given in 
eqn.~\ref{eq-nun4} the evolution of the fundamental constants can also be written as
\begin{equation} \label{eq-wfc}
\frac{\Delta c}{c} = \zeta_c \int_1^a \sqrt{3 \Omega_{\phi}(w+1)}x^{-1}dx
\end{equation}
\citep{cal11,thm12} which shows that whenever $w$ is different from minus one the fundamental
constants are expected to vary making $\mu$ and $\alpha$ $w$ meters in the universe and
excellent discriminators between static and dynamic dark energy.  The beta function, however,
provides a much simpler method for predicting the evolution of the constants as a function of
the cosmology and the dark energy potential.

Since the proton to electron mass ratio $\mu$ has the most reliable and tightest restriction on its
temporal variance it is used as the example in this discussion.  The discussion for the fine
structure constant $\alpha$ is for the most part exactly the same except  for the substitution of
$\zeta_{\alpha}$ instead of $\zeta_{\mu}$ in eqn.~\ref{eq-dx} or~\ref{eq-wfc}.  In both cases
the coupling $\zeta_{\mu,\alpha}$ is considered a constant.  The evolution of $\mu$ as a function
of the scale factor is  simply
\begin{equation} \label{eq-dmu}
\frac{\Delta \mu}{\mu} =\zeta_{\mu} (\sqrt{-2 \beta_p ln(a) + \phi_0^2}-\phi_0), \hspace{0.5cm}  \zeta_{\mu}( \sqrt{2 \beta_i ln(a) + \phi_0^2}-\phi_0)
\end{equation}
for the power law, $\beta_p$ or the inverse power law, $\beta_i$ dark energy potentials.

\begin{figure}
\scalebox{.6}{\includegraphics{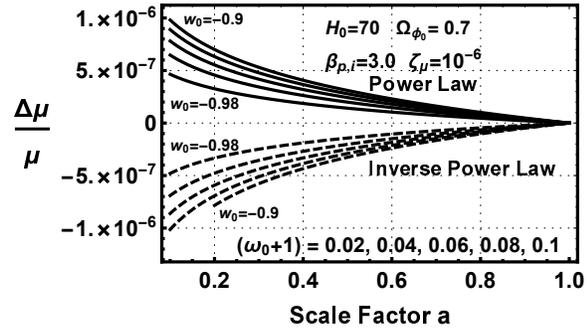}}  
\caption{The evolution of $\frac{\Delta \mu}{\mu}$ for the five different values of $w_0$ with
$\beta_{p,i}$ set to 3.}
\label{fig-dmu}
\end{figure}

\begin{figure}
\scalebox{.6}{\includegraphics{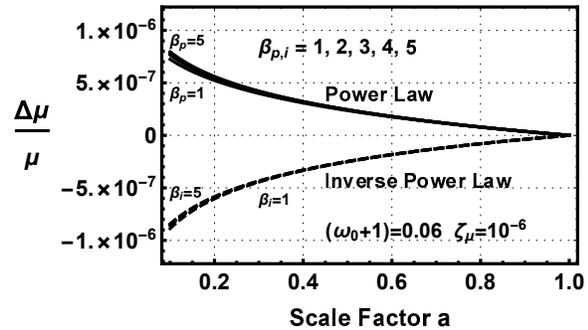}}  
\caption{The evolution of $\frac{\Delta \mu}{\mu}$ for the five different values of $\beta_{p,i}$ 
with $w_0$ set to -0.94}
\label{fig-dmub}
\end{figure}

As an example $\zeta_{\mu}$ is set to $10^{-6}$ and $\beta_{p,i}$ to 3.0.  Figure~\ref{fig-dmu}
shows the evolution for both the power law (solid lines) and the inverse power law (dashed)
line cases.  Since the coupling constant can be either positive or negative the sign of
$\frac{\Delta \mu}{\mu}$ is not a discriminator between the power and inverse power law
potentials unless the sign of the coupling is somehow determined.
The sensitivity to $w_0$ is evident in the figure.  Figure~\ref{fig-dmub} shows the evolution
of $\frac{\Delta \mu}{\mu}$ with $w_0$ held constant at -0.94 with the five different values of
$\beta_{p,i}$. 
As expected the evolution of $\frac{\Delta \mu}{\mu}$ is largely insensitive to $\beta_{p,i}$
since it is proportional to $\Delta \phi$ which is also largely independent of the power of the 
power laws as shown in fig.~\ref{fig-delphi}.  

\subsubsection{Observational Constraints on  $\frac{\Delta \mu}{\mu}$} \label{ss-muob}
\begin{figure}
\scalebox{.6}{\includegraphics{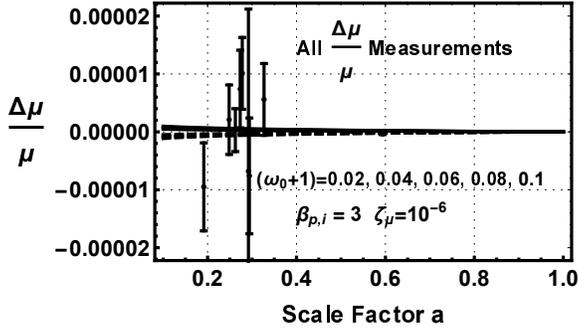}}  
\caption{All of the observational constraints on  $\frac{\Delta \mu}{\mu}$ with the evolution
of $\frac{\Delta \mu}{\mu}$ from fig.~\ref{fig-dmu} superimposed. The radio constraints
at redshifts less than one are not visible at the resolution of this figure.}
\label{fig-alle}
\end{figure}

Observations of molecular absorption lines from cold gas along the line of sight to distant
quasars provide the constraints on $\frac{\Delta \mu}{\mu}$.  Changes in $\mu$ alter the
energy levels of molecules according to the quantum numbers of the upper and lower states
of the transition \citep{thm75} changing the wavelengths of the transitions in a manner that
can not be mimicked by a redshift.  The majority of constraints arise from the observation
of molecular hydrogen absorption lines of the Lyman and Werner bands at redshifts greater
than 2.  More recently radio observations of methanol and ammonia absorption lines at redshifts
less than one have provided more stringent constraints.  The tightest constraints come from
methanol lines in the spectrum of PKS1830-211 at a redshift of 0.88582 by \citet{bag13}
and \citet{kan15} finding  $\frac{\Delta \mu}{\mu} = (-2.9 \pm 5.7)\times 10^{-8}$.
Concerns about common lines of sight has raised the $1\sigma$ error to $\pm10^{-7}$
which will be used here.  Figure~\ref{fig-alle} shows all of the measurements to date.
All of the measurements at redshifts greater than one are optical observations of molecular
hydrogen redshifted into the visible region.  The radio constraints at redshifts less than one
are not visible at the scale of this plot.

\begin{figure}
\scalebox{.6}{\includegraphics{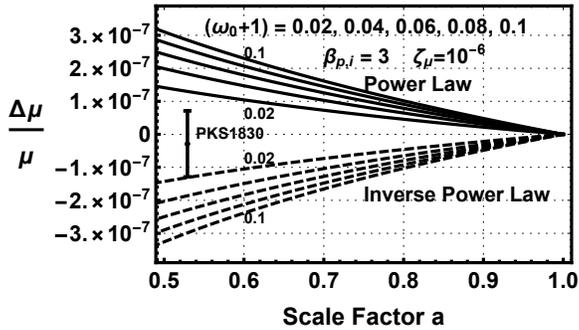}}  
\caption{The radio observational constraints on $\frac{\Delta \mu}{\mu}$ with the evolutionary
tracks from fig.~\ref{fig-dmu}.  For $\beta_{p,i} = 3.0$ only the track with $w_0 = -0.98$
satisfies the constraint at the $1\sigma$ level with $\zeta_{\mu} = 10^{-6}$.}
\label{fig-rad}
\end{figure}

The PKS1830-211 constraint is shown in fig.~\ref{fig-rad} at expanded scale to make the
constraint visible.  
For $\beta_{p,i}=3.0$ and $\zeta_{\mu}=10^{-6}$ the constraint requires $(w_0+1)$ to be 0.02 or
less and is of course consistent with the $\Lambda$CDM value of zero.  If the error bar
was centered on zero then $(w_0+1)$ would have to be less than 0.02.

\begin{figure}
\scalebox{.6}{\includegraphics{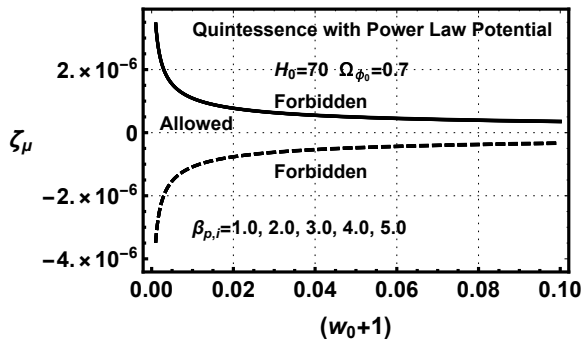}}  
\caption{The allowed and forbidden regions in the $\zeta_{\mu}$ vs $(w_0=1)$ plane
imposed by the limit on $\frac{\Delta \mu}{\mu}$ shown in fig.~\ref{fig-rad}.}
\label{fig-fa}
\end{figure}

Any constraint on $\frac{\Delta \mu}{\mu}$ or $\frac{\Delta \alpha}{\alpha}$ can be met
by either adjusting a cosmological parameter $(w_0+1)$ or a particle physics parameter
$\zeta_{\mu,\alpha}$, therefore, the observations constrain a two dimensional space. The
observations define allowed and forbidden areas in the $\zeta_{\mu,\alpha}$, $(w_0+1)$
parameter space.  Figure~\ref{fig-fa} shows the allowed and forbidden areas defined by
the $\frac{\Delta \mu}{\mu}$ constraint.
More stringent observational bounds on $(w_0+1)$ could place currently allowed regions 
into the forbidden region.  The only point on the diagram consistent with $\Lambda$CDM 
and the standard model is the origin.

The coupling constants $\zeta_{\mu,\alpha}$ are really couplings to the particle physics
parameters, the Quantum Chromodynamic scale, $\Lambda_{QCD}$, the Higgs Vacuum
Expectation Value, $\nu$, and the Yukawa couplings, $h$ \citep{coc07,thm17}.  The fractional
variations, $\frac{\Delta \mu}{\mu}$ and $\frac{\Delta \alpha}{\alpha}$ are two different 
functions of the fractional variations of  $\Lambda_{QCD}$, $\nu$ and $h$.  The combined
limits on the fractional variation of $\mu$ and $\alpha$ then place limits on
$\frac{\Delta \Lambda_{QCD}}{\Lambda_{QCD}}<7.9\times 10^{-5}$ and the sum 
$(\frac{\Delta \nu}{\nu} +\frac{\Delta h}{h})<8.0\times 10^{-5}$ that can not be duplicated 
by laboratory measurements \citep{thm17}.

\section{Relevant but not Directly Observable Parameters} \label{s-rp}
There are several cosmological parameters that are relevant but not directly observable.
Here three parameters, the time derivative of the scalar field, the dark energy density,
and the dark energy pressure, are calculated as functions of the scale factor $a$.

\subsection{The Evolution of the Time Derivative of the Scalar} \label{ss-phidot}
The time derivative of the scalar $\phi$ is an important cosmological parameter that 
appears in both the dark energy pressure and density equations. Since the beta function 
is the derivative of the scalar with respect to the natural log of the scale factor the time 
derivative of the scalar is simply the Hubble parameter times the beta function.
\begin{equation} \label{eq-pdot}
\dot{\phi}= a\frac{d \phi}{da}\frac{\dot{a}}{a}=\beta H =-\frac{1}{2}\beta W
\end{equation}

\begin{figure}
\scalebox{.6}{\includegraphics{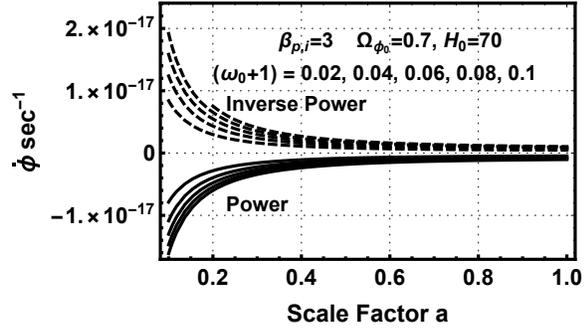}} 
\caption{The time derivative of the scalar. The $(w_0+1)=0.1$ case has the most evolution
and $(w_0+1)=0.02$ has the least evolution for both the power (solid lines) and inverse power 
(dashed lines) law potentials.}
\label{fig-pdot}
\end{figure}

Figure~\ref{fig-pdot}
shows the evolution of $\dot{\phi}$ with respect to the scale factor $a$.  Since $H(a)$ is
essentially invariant to either the power of the dark energy potential or the value of $w_0$
the dependence on $w_0$ is entirely due to the beta functions' dependence on $\beta_{p,i}$
and $w_0$.  Figures~\ref{fig-betaw} and~\ref{fig-betab} show that the main dependence
is on $w_0$ as opposed to $\beta_{p,i}$.

An analytic expression for $\dot{\phi}$ is obtained by multiplying either eqn.~\ref{eq-awa}
or~\ref{eq-awai} by $(-\frac{1}{2})\beta(\phi)$ with $\beta(\phi)$ given by the appropriate
functions in eqns.~\ref{eq-betap} and~\ref{eq-phi}.  An alternative is to use the functions
in eqns.~\ref{eq-hcomp} and~\ref{eq-hcomi} for $H(a)$ resulting in
\begin{equation} \label{eq-pdp}
\dot{\phi_p}(a)=\frac{\beta_p}{2 \phi_p(a)}W_p(a)
\end{equation}
for the power law potential and
\begin{equation} \label{eq-pdi}
\dot{\phi_i}(a)= \frac{-\beta_i}{2\phi_i(a)}W_i(a)
\end{equation}
for the inverse power law.  In eqns~\ref{eq-pdp} and~\ref{eq-pdi} the approximate forms
of $H(a)$ can also be used.  Using the Gamma function forms is slightly more accurate.

It is obvious from fig.~\ref{fig-pdot} that although there is significant early time evolution of
$\dot{\phi}$ the late time evolution is a slow approach to zero.  This indicates that power and
inverse power law quintessence predicts very small time variations of the fundamental constants
at the present time.  This is a general characteristic of most freezing cosmologies where $w$ is 
initially different from minus one and evolves toward minus one with time.  The power law 
values of $\dot{\phi}$ are negative since the scalar is decreasing while the inverse power law 
values are positive since the scalar is increasing with time for this case.

\subsection{The Evolution of the Dark Energy Density and Pressure} \label{ss-dedp}
From eqn.~\ref{eq-em1} it is clear that 
\begin{equation} \label{eq-rphi}
\rho_{\phi} = 3 H^2 - \rho_m = 3H^2(a) -\frac{\rho_{m_0}}{a^3}
\end{equation}
which is consistent with eqn. 3.8 from \citet{cic17} which gives the total potential with mass
as
\begin{equation} \label{eq-vm}
V=\rho_{\phi} - \frac{1}{2}\dot{\phi}^2=3H^2-\rho_m  - \frac{1}{2}\dot{\phi}^2
\end{equation}

\begin{figure}
\scalebox{.6}{\includegraphics{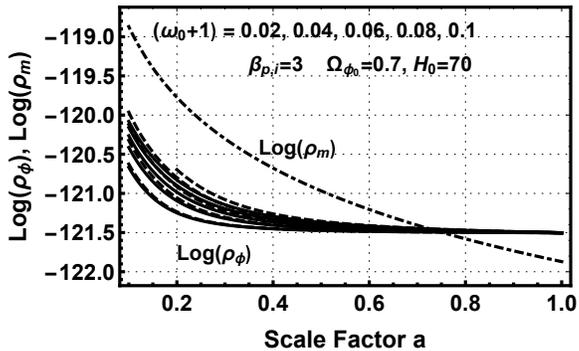}} 
\caption{The power law potential (solid line) and inverse power law (dashed line) dark
energy density values as a function of the scale factor.  In both cases the $w_0 = -0.9$ 
tracks have the highest values and the $w_0=-0.98$ tracks have the lowest values.  The
dashed-dot line is the matter density which decreases below the dark energy density near
a scale factor of 0.75}
\label{fig-aden}
\end{figure}

Figure~\ref{fig-aden} shows the evolution of the densities using the Gamma function
equations~\ref{eq-awa} and~\ref{eq-awai}  to compute $H(a)$.  The matter 
density, shown by the dash dot line, is also plotted to indicate the crossover from matter 
to dark energy dominated evolution.
For values of $(w_0+1)$ close to zero the power and inverse power law plots nearly
overlap but as $(w_0+1)$ diverges from zero the inverse power law cases have slightly
higher densities at scale factors less than 0.5.  All cases converge to the boundary condition
on the density at a scale factor of one.  Most of the evolution of $\rho_{\phi}$ occurs at
scale factor less than 0.3, consistent with the previous plots of the evolution of $(w+1)$ ib
fig.~\ref{fig-ww} and $\dot{\phi}^2$ in fig.~\ref{fig-pdot}.

\begin{figure}
\scalebox{.6}{\includegraphics{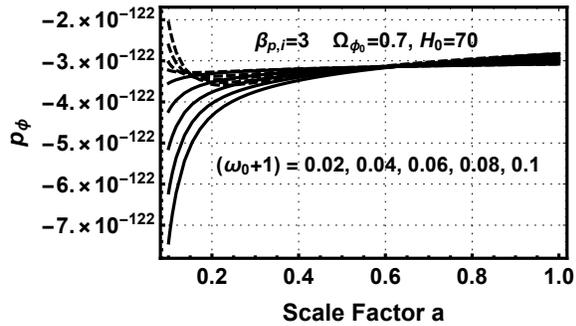}}  
\caption{The dark energy pressure. The $(w_0+1)=0.1$ case has the most evolution
and $(w_0+1)=0.02$ has the least evolution for both the power (solid lines) and inverse power 
(dashed lines) law potentials.}
\label{fig-allp}
\end{figure}

The dark energy pressure is also given in eqns.~\ref{eq-rhop}.  
\begin{equation} \label{eq-dep}
p_{\phi}(a) = \dot{\phi}^2 -3H(a)+\frac{\rho_{m_0}}{a^3}
\end{equation}
Figure~\ref{fig-allp} shows the evolution of the dark energy pressure.
As was the case with the dark energy density the inverse power law dark energy potential case
has more evolution than the power law case, particularly for $w_0$ values further from minus
one.  Both the dark energy density and the dark energy potential are not significantly dependent
on the power $\beta_{p,i}$ for a given value of  $w_0$.

\section{Summary} \label{s-sum}
The beta function formalism is demonstrated using the example of the quintessence 
cosmology with power and inverse power law dark energy potentials.  Simple beta
functions were found, $\beta(\phi) = \pm\frac{\beta_{p,i}}{\phi}$ where 
$\beta_{p,i}$ is a constant equal to the power.  The minus sign 
applies to the power law, $p$, and the positive sign to the inverse power law, $i$. From the
beta functions the scalar $\phi$, as a function of the scale factor $a$ is calculated with a
boundary condition supplied by the current value of the dark energy equation of state $w$.  This 
provides an easy transition from functions of the generally unobservable scalar $\phi$ to functions
of the easily observable scale factor $a=\frac{1}{1+z}$. Beta potentials are produced that 
reproduce the model dark energy potentials to better than one percent.  These potentials 
produce actions that accurately represent the actions with the model potentials.  The extra
beta function combined with the quintessence equations for the dark energy pressure and 
density plus the usual cosmological equations provide the means to calculate an analytic
function for the super potential, $W=-\frac{1}{2}H$ where H is the Hubble parameter.

The super potential automatically provides the Hubble parameter as a function of the scale
factor.  It is found that the Hubble parameter is essentially insensitive to the power
of the potential or $w_0$ and includes the $\beta_{p,i} =0$ case which corresponds to the 
$\Lambda$CDM cosmology.  This demonstrates that the Hubble parameter is not a
good indicator to discriminate between static and dynamical dark energy.  It is confirmed
that the transition from matter dominated to dark energy dominated epochs occurs
at the proper time and that the evolution of the Hubble parameter matches a randomly
selected current list of $H(z)$ measurements.  The measurements also provide a
best fit value of $H_0$ for the selected data set.

Additional observable parameters, the dark energy equation of state, and the variation
of the fundamental constants $\mu$ and $\alpha$ in a rolling scalar field are calculated.
The limits on the variation of the constants imposes allowed and forbidden regions in the
two dimensional $w+1$, $\zeta_{\mu}$ plane in a balance between cosmological and 
elementary particle physics parameters.  Analytic expressions for three not directly 
observable parameters, $\dot{\phi}$, $\rho_{\phi}$ and $p_{\phi}$ are also calculated.
It is generally noted that the parameter evolution is more sensitive to the current value
of the dark energy equation of state $w_0$ than the power of the potentials $\beta_{p,i}$.

This work demonstrates of the power of the beta function formalism to produce accurate 
predictions for comparison with observation.  The formalism is expandable to other forms 
of the dark energy potential and other cosmologies which will be the subject of future work.

\label{lastpage}
\end{document}